\RequirePackage{fixltx2e}
\documentclass[twoside,journal, letterpaper, final, twosided]{IEEEtran}
\usepackage[T1]{fontenc}
\usepackage[latin9]{inputenc}
\usepackage{amstext}
\usepackage{graphicx}

\makeatletter

\newcommand{\lyxdot}{.}

\setkeys{Gin}{width=\linewidth}
\usepackage{cite}

\long\def\@makecaption#1#2{\ifx\@captype\@IEEEtablestring%
\footnotesize\begin{center}{\normalfont\footnotesize #1}\\
{\normalfont\footnotesize\scshape #2}\end{center}%
\@IEEEtablecaptionsepspace
\else
\@IEEEfigurecaptionsepspace
\setbox\@tempboxa\hbox{\normalfont\footnotesize {#1.}~~ #2}%
\ifdim \wd\@tempboxa >\hsize%
\setbox\@tempboxa\hbox{\normalfont\footnotesize {#1.}~~ }%
\parbox[t]{\hsize}{\normalfont\footnotesize \noindent\unhbox\@tempboxa#2}%
\else
\hbox to\hsize{\normalfont\footnotesize\hfil\box\@tempboxa\hfil}\fi\fi}
\ifCLASSOPTIONcompsoc
  \usepackage[caption=false,font=normalsize,labelfont=sf,textfont=sf]{subfig}
\else
  \usepackage[caption=false,font=footnotesize]{subfig}
\fi

\usepackage{amsmath}
\DeclareMathOperator*{\argmax}{arg\,max}

\usepackage{dblfloatfix}

\@ifundefined{showcaptionsetup}{}{%
 \PassOptionsToPackage{caption=false}{subfig}}
\usepackage{subfig}
\makeatother

\begin{document}

\title{Bio-inspired Filter Banks for Frequency Recognition of SSVEP-based
Brain-computer Interfaces}

\author{Ali Fatih Demir, Student Member, IEEE, Huseyin Arslan, Fellow, IEEE,
Ismail Uysal, Member, IEEE\vspace{-0mm}
\thanks{Manuscript received September 30, 2018; revised February
09, 2019, May 28, 2019, August 20, 2019, and September 20, 2019; accepted
October 09, 2019. This work was supported by the Southeastern Center
for Electrical Engineering Education (SCEEE) Research Initiation Grant
\# SCEEE16-003.} \thanks{Ali Fatih Demir and Ismail Uysal are with
the Department of Electrical Engineering, University of South Florida,
Tampa, FL 33620 USA (e-mail: afdemir@mail.usf.edu, iuysal@usf.edu).}\thanks{Huseyin
Arslan is with the Department of Electrical Engineering, University
of South Florida, Tampa, FL 33620 USA and also with the College of
Engineering, Istanbul Medipol University, Istanbul 34810, Turkey (e-mail:
arslan@usf.edu).}}
\maketitle
\begin{abstract}
Brain-computer interfaces (BCIs) and their associated technologies
have the potential to shape future forms of communication, control,
and security. Specifically, the steady-state visual evoked potential
(SSVEP) based BCIs have the advantages of better recognition accuracy,
and higher information transfer rate (ITR) compared to other BCI modalities.
To fully exploit the capabilities of such devices, it is necessary
to understand the underlying biological features of SSVEPs and design
the system considering their inherent characteristics. This paper
introduces bio-inspired filter banks (BIFBs) for improved SSVEP frequency
recognition. SSVEPs are frequency selective, subject-specific, and
their power gets weaker as the frequency of the visual stimuli increases.
Therefore, the gain and bandwidth of the filters are designed and
tuned based on these characteristics while also incorporating harmonic
SSVEP responses. The BIFBs are utilized in the feature extraction
stage to increase the separability of classes. This method not only
improves the recognition accuracy but also increases the total number
of available commands in a BCI system by allowing the use of stimuli
frequencies that elicit weak SSVEP responses. The BIFBs are promising
particularly in the high-frequency band, which causes less visual
fatigue. Hence, the proposed approach might enhance user comfort as
well. The BIFB method is tested on two online benchmark datasets and
outperforms the compared methods. The results show the potential of
bio-inspired design, and the findings will be extended by including
further SSVEP characteristics for future SSVEP based BCIs.
\end{abstract}


\markboth{IEEE ACCESS}{Demir \MakeLowercase{\textit{et al.}}:Bio-inspired Filter Banks for Frequency Recognition of SSVEP-based Brain-computer Interfaces}

\section{Introduction}

\begin{figure}
\centering\includegraphics[width=1\columnwidth]{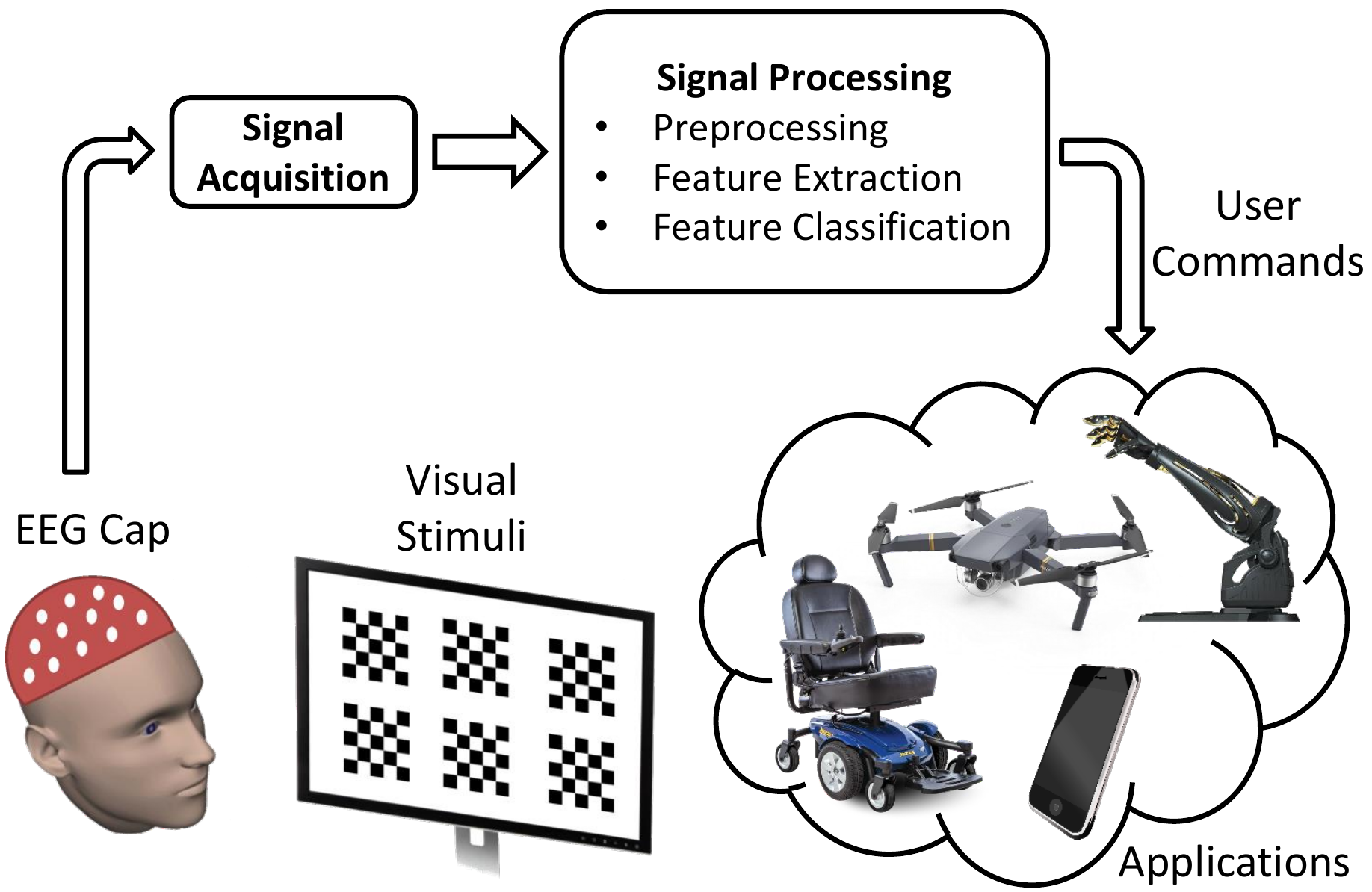}
\centering{}\caption{Functional model of an SSVEP-based BCI.\label{fig:BCI}}
\end{figure}

\IEEEPARstart{S}{cientific} advances in neuroscience and biomedical
engineering enabled a direct communication channel between the human
brain and a computer. The electrical activity in the brain that is
produced by neuronal post-synaptic membrane polarity changes can be
monitored to detect the user\textquoteright s intentions \cite{shih2012}.
A brain-computer interface (BCI) \cite{vidal1973} analyzes the brain
signals and translates them into commands for external devices such
as a speller device, wheelchair, robotic arm, or a drone (Fig. \ref{fig:BCI}).
Since BCIs utilize the signals generated by the central nervous system,
the primary target of this technology is people with severe neuromuscular
disorders (e.g., amyotrophic lateral sclerosis, brain-stem stroke,
spinal cord injury, and cerebral palsy). However, advanced BCI systems
serve healthy people as well by providing an alternative way of communication,
control, and security \cite{wolpaw2002,gao2014,valenzuela2017}. Hence,
these systems have evolved to be a promising part of the body area
network \cite{schirner2013,movassaghi2014,demir2016invivo,demir2017anatomical,wang2018}.

While there exist multiple approaches to measure brain activity, electroencephalography
(EEG) is widely used in BCI applications because of its high temporal
resolution, which is essential for BCIs to work as real-time systems
\cite{vialatte2010}. In addition, EEG devices are inexpensive and
portable. Various EEG signals could serve to drive BCIs. For example,
a distinctive oscillation pattern in EEG is observed when a sensory
stimulus such as visual or auditory is presented to a human. These
oscillations are called as evoked potentials (EPs), and they disappear
after a short period. If the stimulus is repeated at a regular rate,
the EPs do not have time to decay, and it causes a periodic response
which is called as steady-state evoked potentials \cite{paris2017}.
More specifically, a periodic visual stimulus with a repetition rate
higher than 6 Hz elicits steady-state visual evoked potentials (SSVEPs)
which are more prominent in the occipital region of the brain \cite{herrmann2001,wolpaw2012}.
The targets that evoke SSVEPs are encoded in various ways \cite{gao2014,zhu2010},
and the users make a selection by shifting their attention to the
desired target in SSVEP based BCIs. Among other BCI modalities which
depend on other EEG signals (e.g., slow cortical potentials, sensorimotor
rhythms, and event-related potentials), SSVEP based BCIs have the
advantage of high information transfer rate (ITR) and short training
duration to operate the device \cite{wang2006}.

\begin{figure}
\centering\includegraphics[width=1\columnwidth]{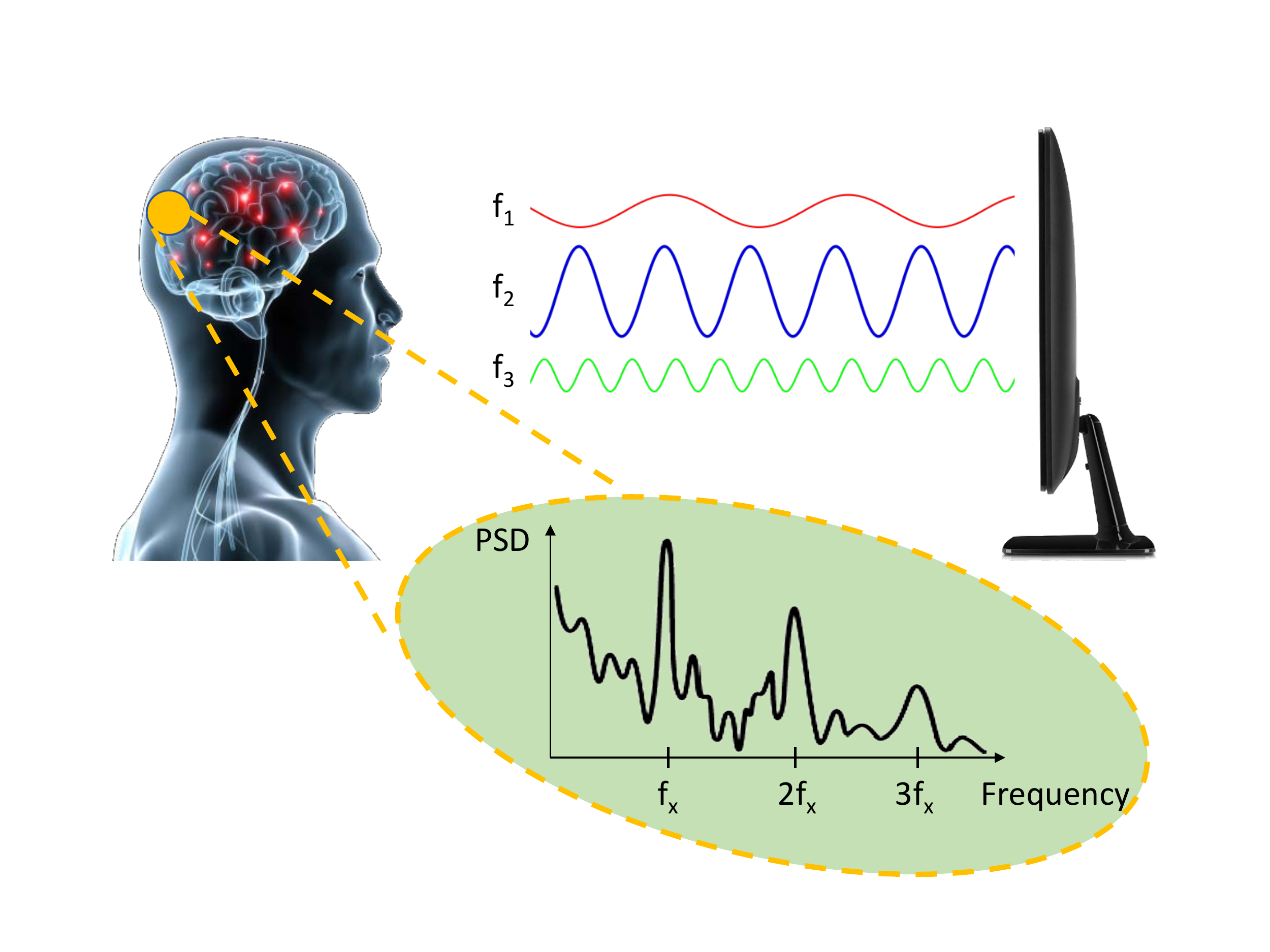}\caption{SSVEP response to frequency-coded stimuli at the occipital region
of the brain.\label{fig:SSVEP}}

\end{figure}

SSVEPs are sinusoidal-like waveforms, and they appear at the same
fundamental frequency of the driving stimulus and its harmonics (Fig.
\ref{fig:SSVEP}) \cite{herrmann2001}. However, spontaneous oscillations
(i.e., background activity), which are not related to the stimulation,
exist in the EEG recordings as well and a robust recognition algorithm
is required to build a reliable BCI system. Numerous methods have
been proposed for SSVEP recognition in the last decade \cite{wang2006,lin2007,bakardjian2010,zhang2013,chen2015,wang2016,bittencourt2018}.
Power spectral density analysis (PSDA) is a typical approach since
the distinctive features of SSVEPs are observed in the frequency domain
\cite{wang2006}. However, PSDA is susceptible to noise, and long
durations are needed to increase the signal to noise ratio (SNR).
A multivariable statistical method, namely canonical correlation analysis
(CCA) \cite{lin2007,zhang2013} exploits the multiple channel covariance
information to enhance SNR and provide a better recognition accuracy
compared to PSDA. Simple implementation, high robustness, and better
ITR performance have made CCA attractive in SSVEP recognition research.
On the other hand, CCA is not efficient to extract the discriminative
information embedded in the harmonic components of SSVEPs, and filter-bank
canonical correlation analysis (FBCCA) \cite{chen2015} is proposed
to handle this issue. Although FBCCA captures the distinct spectral
properties of multiple harmonic frequencies successfully, it neglects
any correlation information between SSVEP responses at different frequencies
\cite{wang2016}. Furthermore, this approach disregards the frequency
selective nature of SSVEPs due to the utilization of wide-band filters
which cover the whole stimuli bandwidth. 

\begin{figure}
\centering\includegraphics[width=1\columnwidth]{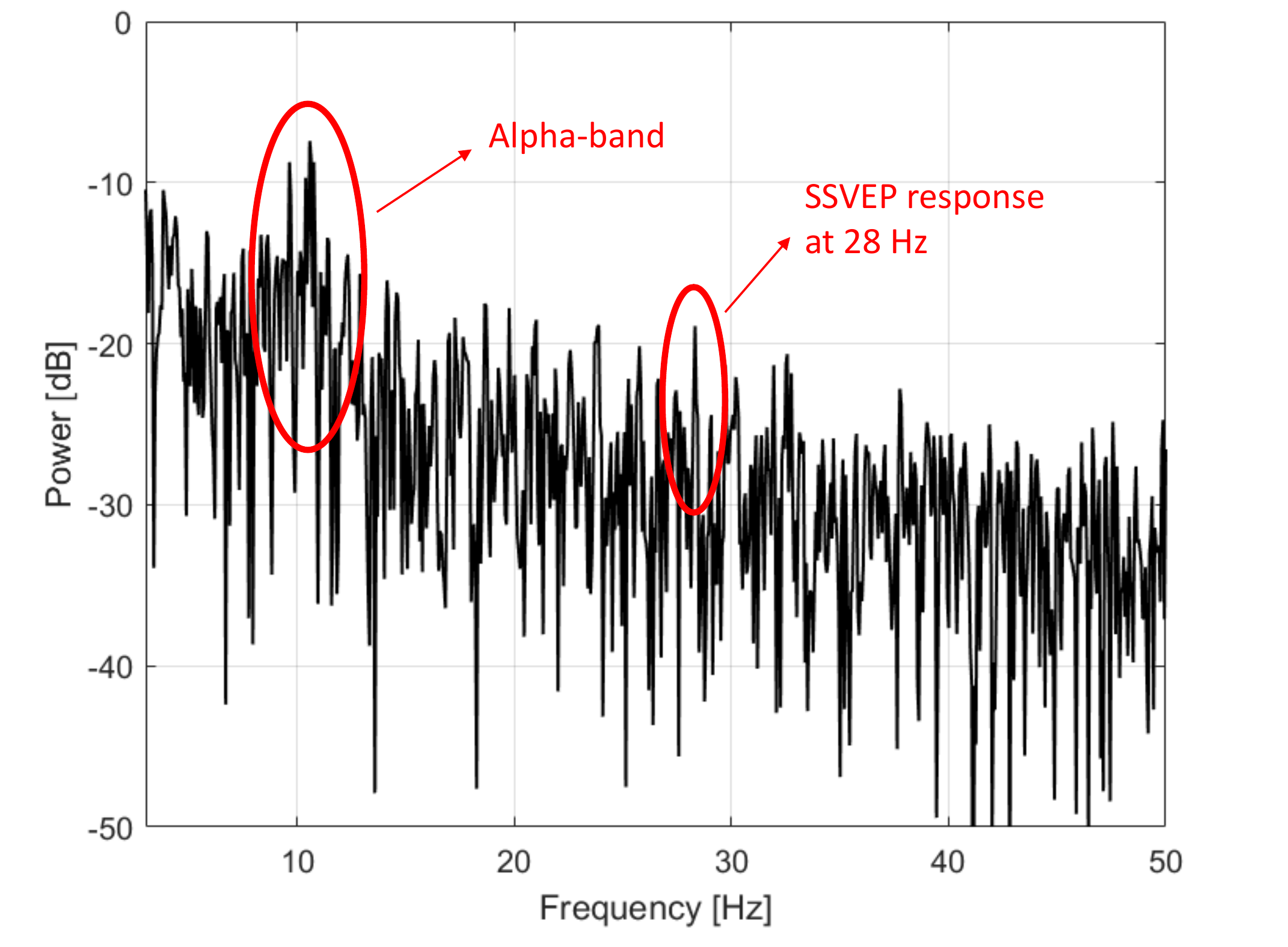}\caption{The PSD of EEG signal when a visual stimulus at 28 Hz is presented
to a participant. \label{fig:WeakSSVEP}}
\end{figure}

To fully exploit and further increase the potential of SSVEP based
BCIs, it is necessary to employ an accurate SSVEP model in the recognition
algorithm. For example, the inclusion of SSVEP harmonics in a recognition
algorithm improves the accuracy \cite{muller-putz2005} since the
spontaneous EEG oscillations typically do not present any harmonic
components \cite{birca2006}. Also, the subject-specific nature of
SSVEPs is handled by an individualized parameter optimization and
calibration (e.g., time-window duration, number of harmonics considered,
and electrode location) \cite{wang2006,lin2007}. Moreover, the SSVEP
response is frequency selective, and its power gets weaker as the
frequency of the stimuli increases \cite{vialatte2010,herrmann2001,bakardjian2010,zhu2010}.
Although the power of EEG background activity decreases as well with
the increase in frequency (approximately with a $1/f$ behavior \cite{paris2017}),
the resultant SNR is still considerably low at high frequencies. Hence,
a visual stimulus at a high frequency can almost be indistinguishable
in the presence of noise as shown in Fig.~\ref{fig:WeakSSVEP}. This
inherent feature not only results in a lower recognition accuracy
but also causes exclusion of the stimulus frequencies that evoke weak
SSVEP response and decreases the total number of available commands
in a BCI system.

This paper introduces bio-inspired filter banks (BIFBs) for improved
SSVEP frequency recognition. The BIFBs are designed considering the
inherent biological characteristics of SSVEPs, namely frequency selectivity,
subject specificity, and harmonic SSVEP responses. They are utilized
in the feature extraction stage to increase the separability of classes.
The proposed approach is tested on datasets available online, and
its performance is compared with the performances of various SSVEP
frequency recognition methods. The preliminary results without an
elaborate classification algorithm or a cross-validation procedure
were presented in \cite{demir2016bifb}. Also, a fair performance
comparison with the utilization of unit filters is provided to validate
the effectiveness of the proposed filter bank design in this study.
The results show a notable ITR improvement with the bio-inspired design
and highlight the promising potential of BIFBs in the high-frequency
band, which causes less visual fatigue. Hence, the proposed method
leads to more reliable, efficient, and user-friendly SSVEP-based BCI
systems.

This article is structured as follows. Section \ref{sec:II} describes
the performance metrics, evaluation methodology, and datasets. The
proposed method is explained in detail, along with the comparison
methods. Section \ref{sec:III} presents the performance of the SSVEP
recognition algorithms and provides a thorough analysis of the results.
Finally, Section \ref{sec:IV} summarizes the contributions and addresses
future research directions.

\begin{table*}
\caption{Overview of the SSVEP datasets \label{tab:Datasets}}

\centering\includegraphics[width=1.6\columnwidth]{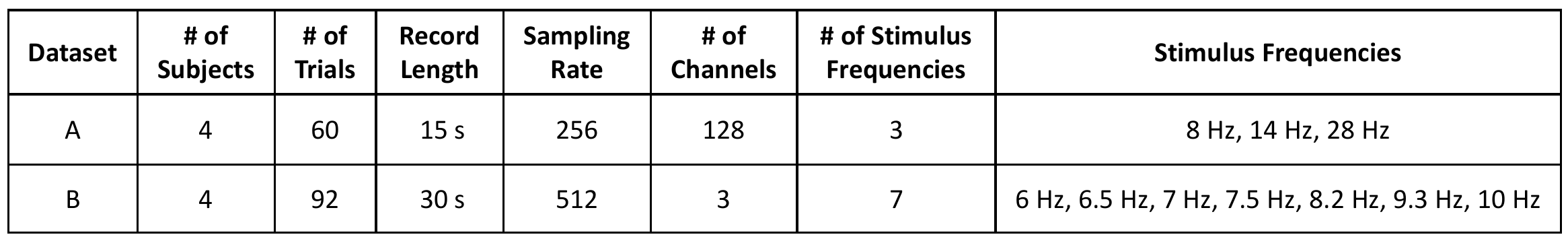}
\end{table*}

\section{Methods and Materials\label{sec:II}}

\subsection{Evaluation Metric}

The most common measure to evaluate the performance of a BCI system
is ITR \cite{wolpaw2002}, which can be expressed in bits/minutes
as follows:

\begin{equation}
ITR=s\left[log_{2}(K+\delta\,log_{2}\delta+(1-\delta)\,log_{2}\left(\frac{1-\delta}{K-1}\right)\right]\label{eq:ITR}
\end{equation}
where $K$ stands for the number of equiprobable commands, $s$ denotes
the commands performed per minute, and $\delta$ represents the accuracy
of target recognition. In general, the BCIs with high ITR have a large
number of commands. However, $K$ is fixed in these datasets, and
the ITR can be boosted with the joint optimization of $s$ and $\delta$.
Also, a threshold can be set either on $s$ or $\delta$ based on
user comfort.

\subsection{Datasets and Pre-processing}

Two publicly-available datasets are utilized in this study to test
the proposed method. Dataset-A \cite{bakardjian2010} consists of
EEG recordings belong to four healthy subjects with normal or corrected
to normal vision. Small reversing black and white checkerboards were
presented to the participants sequentially (i.e., one stimulus at
a time) at three different frequencies (8 Hz, 14 Hz, and 28 Hz) during
the recordings. The brain signal acquisition was performed at a sampling
rate of 256 Hz with 128 active electrodes using the ABC layout standard
\footnote{https://www.biosemi.com/headcap.htm} for electrode placement.
The EEG recordings were re-referenced using the central Cz electrode
and band-pass filtered from 6 Hz to 35 Hz. The subjects experienced
a visual stimulus for 15 seconds in each trial. Each unique visual
stimulus was repeated for five times, which corresponds to 60 trials
(4 subjects x 3 stimuli x 5 repetitions) in total. Dataset-B \cite{2013AVI},
which is provided by another research institute, consists of EEG recordings
belong to four healthy subjects as well. A single flickering box that
changes color rapidly from black to white at seven different frequencies
(6 Hz, 6.5 Hz, 7 Hz, 7.5 Hz, 8.2 Hz, 9.3 Hz, and 10 Hz) was used as
the visual stimulus. The brain signal acquisition was performed at
a sampling rate of 512 Hz with three electrodes (Oz, Fpz, Pz) using
the 10-20 layout standard for electrode placement. The EEG recordings
were referenced using the electrode Fz, and an analog notch filter
at 50 Hz was applied to suppress the power-line noise. The subjects
experienced a visual stimulus for 30 seconds in each trial. Each unique
visual stimulus was repeated at least three times with 92 trials in
total. 

An overview of these datasets is provided in Table \ref{tab:Datasets},
and the reader is referred to individual references for a more detailed
description of the datasets. Dataset-A is selected to include a stimulus
at the high-frequency band that evokes weak SSVEP response, whereas
Dataset-B is selected to deal with the frequency selectivity even
in a narrow band.

\subsection{Proposed Method \label{subsec:PropAlg}}

The pre-processed EEG signal from the occipital channel Oz is segmented
with an overlap, and each segment is windowed using a Hamming function
\cite{oppenheim2009}. Afterward, the power spectral density of the
signal is estimated by the following equation:
\begin{figure}[b]
\centering\includegraphics[width=1\columnwidth]{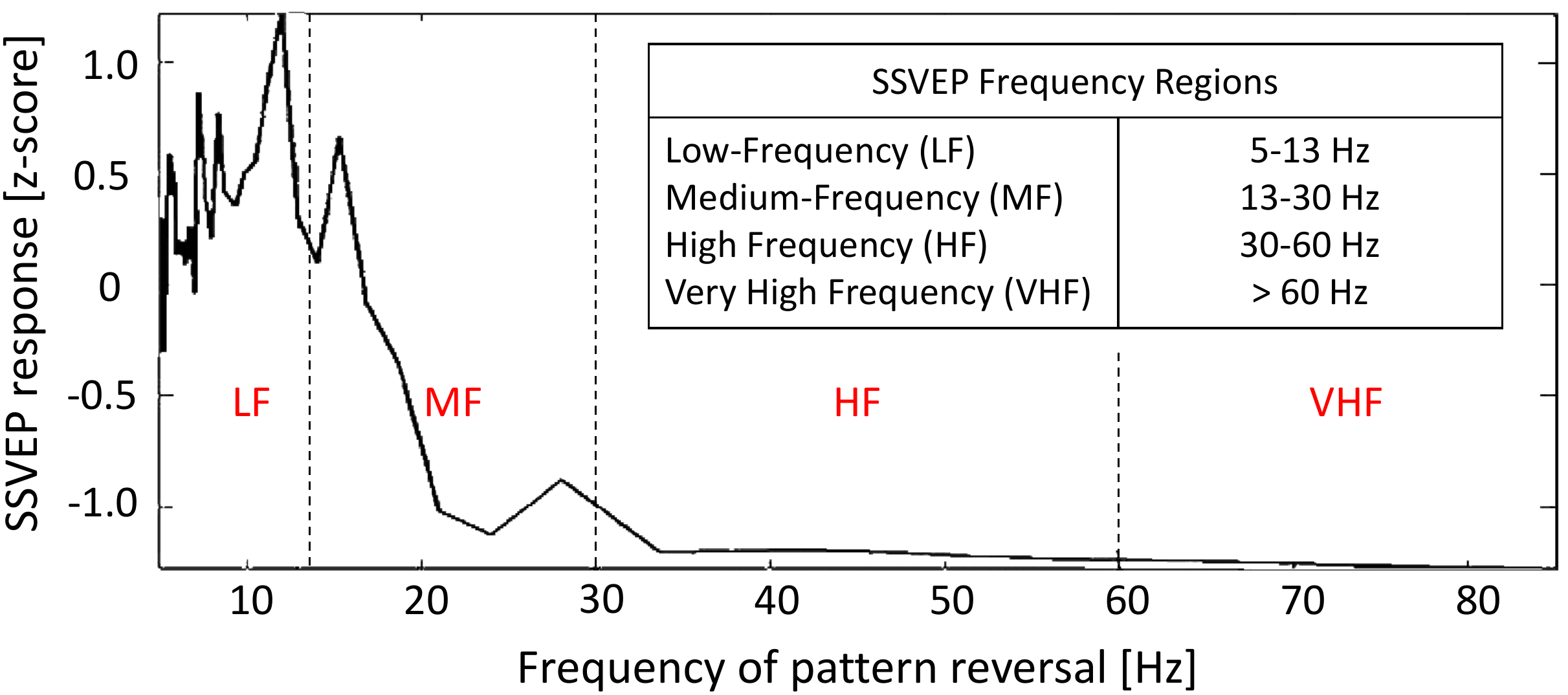}\caption{SSVEP response to pattern reversal stimuli ranging in frequency from
5.1 Hz to 84 Hz \cite{bakardjian2010}. \label{fig:SSVEPresp}}
\end{figure}

\begin{equation}
S_{EEG}[f]=\frac{1}{N}\left|\sum_{n=0}^{N-1}EEG[n]\:w[n]\:e^{-j(\frac{2\pi fn}{N})}\right|^{2}
\end{equation}
where $EEG[n]$ and $w[n]$ represent the discrete EEG signal and
Hamming window function, respectively. The features for SSVEP frequency
recognition are extracted by multiplying $S_{EEG}$ with the frequency
response of BIFBs. The filter banks are designed in such a way that
they capture the inherent biological characteristics of the SSVEPs.
It is known that the SSVEPs are frequency-selective, and their power
gets weaker as the frequency of the visual stimuli increases \cite{vialatte2010,herrmann2001,bakardjian2010,zhu2010}.
Figure \ref{fig:SSVEPresp} presents the average SSVEP response power
to pattern reversal stimuli ranging from 5.1 Hz to 84 Hz \cite{bakardjian2010}.
Especially, the stimuli at the high-frequency bands elicit weak responses
and make the recognition challenging. Consequently, the gain and bandwidth
of the filters are designed considering the frequency-selective nature
of SSVEPs. Assume that there are $K$ target stimulus frequencies
($\tilde{f_{k}})$, where $k=\left\{ 1,...,K\right\} $, in a BCI
system. The array of filters in BIFBs is expressed as follows:

\begin{equation}
H_{BIFB}^{k}[f]:\left\{ \begin{array}{cc}
\frac{f-(\tilde{f_{k}}-BW_{k}/2)}{BW_{k}}\:g_{k}, & (\tilde{f_{k}}-BW_{k}/2)\leq f\leq\tilde{f_{k}}\\
\frac{(\tilde{f_{k}}+BW_{k}/2)-f}{BW_{k}}\:g_{k}, & \tilde{f_{k}}\leq f\leq(\tilde{f_{k}}+BW_{k}/2)\\
0, & otherwise
\end{array}\right.
\end{equation}
where $BW_{k}$ and $g_{k}$ represent the bandwidth and gain of the
$k^{th}$ filter, respectively. Initially, higher bandwidth and gain
are set to frequencies with low SSVEP response power. Subsequently,
these parameters are optimized for individual users in order to counter
the subject-specific nature of SSVEP response \cite{wang2006,lin2007}.
A grid search algorithm performed this hyper-parameter optimization
through a manually specified subset of the hyper-parameter space \cite{bergstra2011}.
It should be noted that the initial parameter guesses considering
the average SSVEP response decrease the computational complexity.
Also, SSVEPs occur at the fundamental frequency of the driving stimulus
and its harmonics, whereas spontaneous EEG oscillations typically
do not present any harmonic components \cite{birca2006}. Accordingly,
filters at the SSVEP harmonic frequencies are included in the filter
bank design (i.e., $H_{BIFB}^{K+1}[f]$ for $2\tilde{f_{1}}$$,...,$
$H_{BIFB}^{K+K}[f]$ for $2\tilde{f_{k}}$) as well to improve the
recognition accuracy as shown in Fig. \ref{fig:SSVEPdesign}. Finally,
the features are extracted using the BIFBs as follows:
\begin{figure}[t]
\centering\includegraphics[width=0.82\columnwidth]{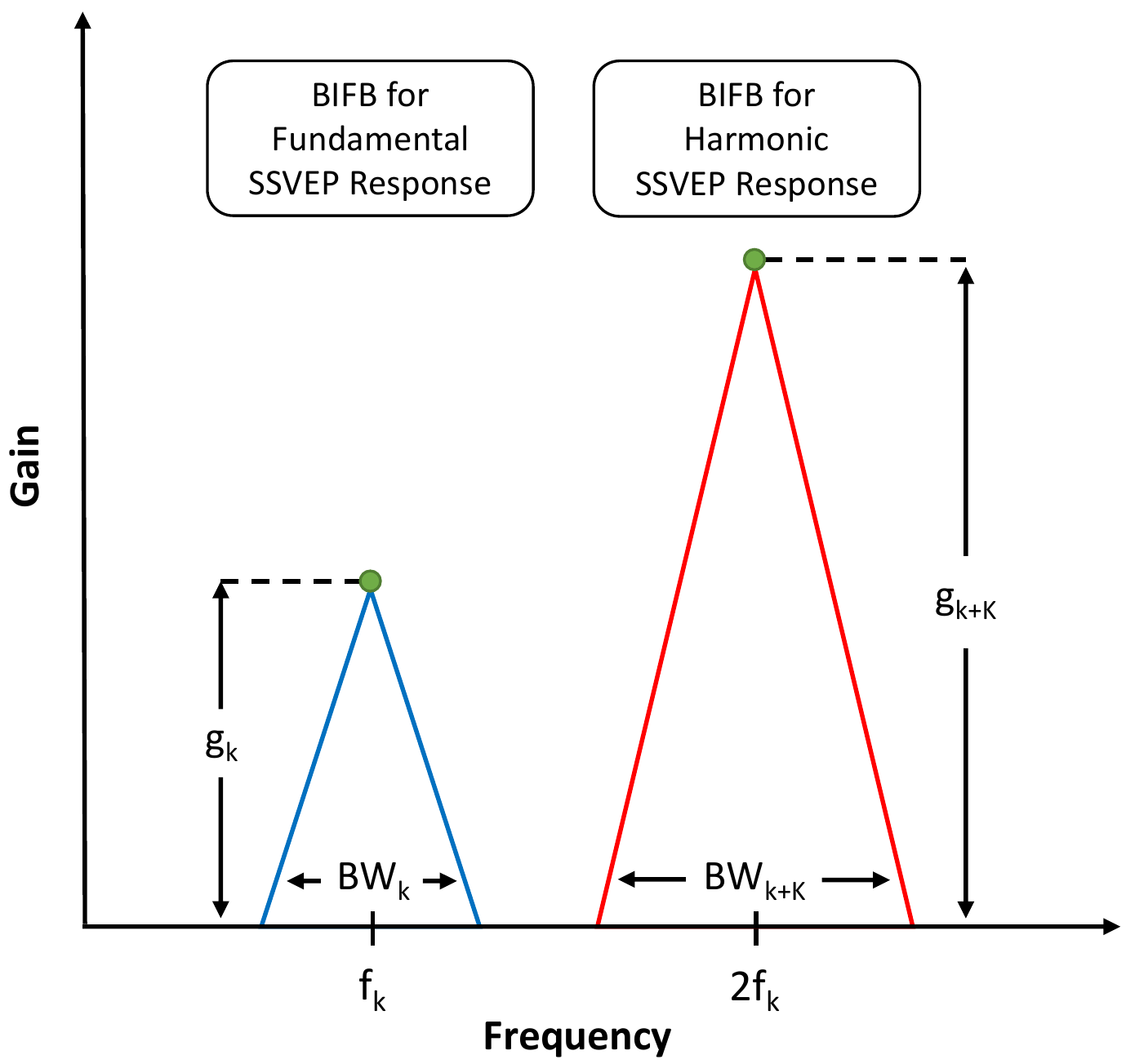}\caption{A bio-inspired filter design to capture SSVEP response at $\tilde{f_{k}}$.
\label{fig:SSVEPdesign}}
\end{figure}
\begin{equation}
\begin{array}{cc}
x_{i}={\displaystyle \sum_{f}S_{EEG}[f]\,H_{BIFB}^{i}[f]} & i=1,...,2K\end{array}
\end{equation}
where $x_{i}$ represents the elements of feature vector $X$.

The extracted features for SSVEP recognition are classified with a
logistic regression model using the one-vs-all strategy. Assume $K$
classes where each class represents a target stimulus frequency. The
hypothesis function predicts whether a given input belongs to $k^{th}$
class or not, and it is formulated by the following equation: 
\begin{figure*}[t]
\subfloat[\label{fig:BIFB-A}]{\includegraphics[width=1\columnwidth]{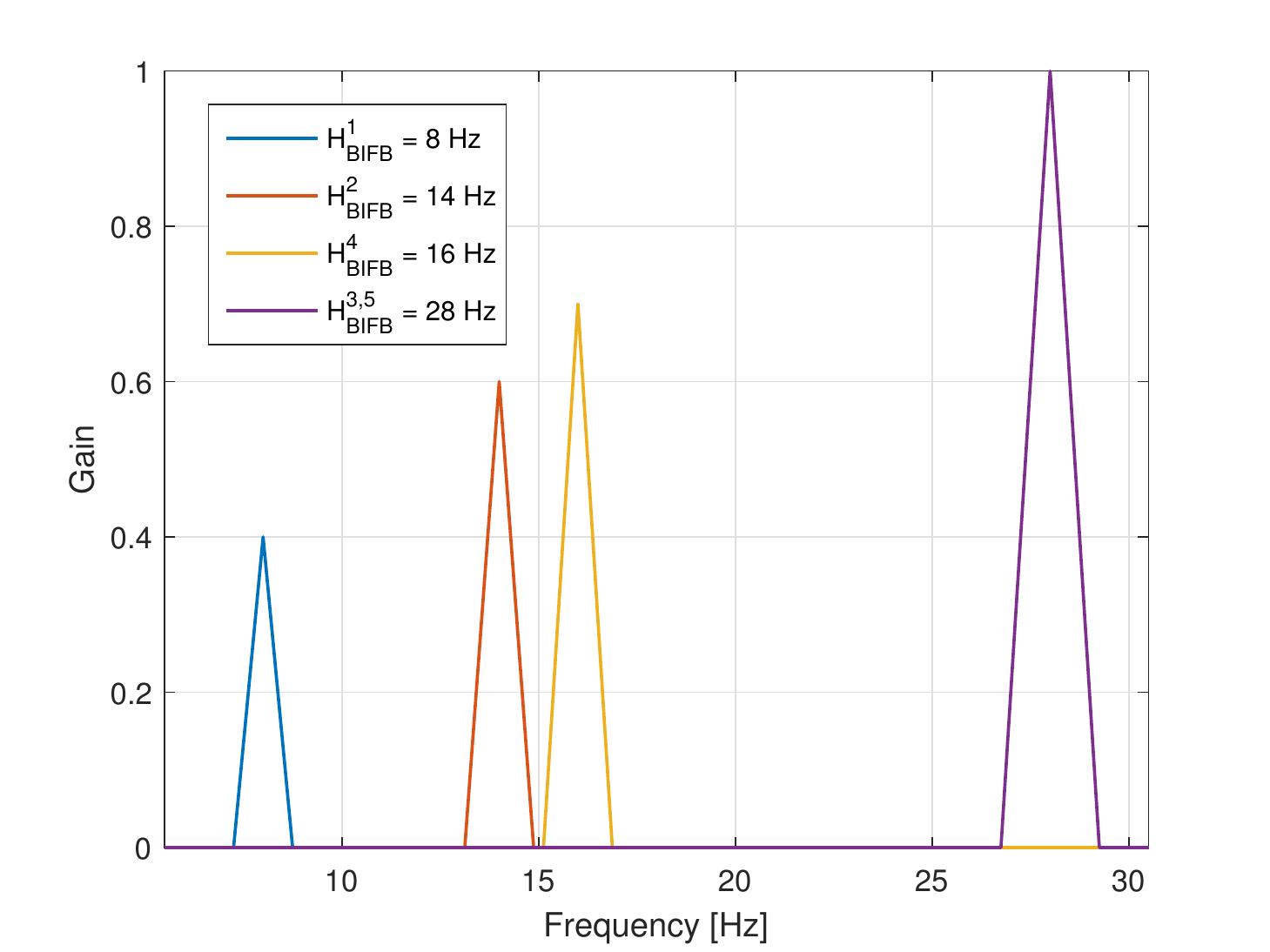}

}\subfloat[\label{fig:BIFB-B}]{\includegraphics[width=1\columnwidth]{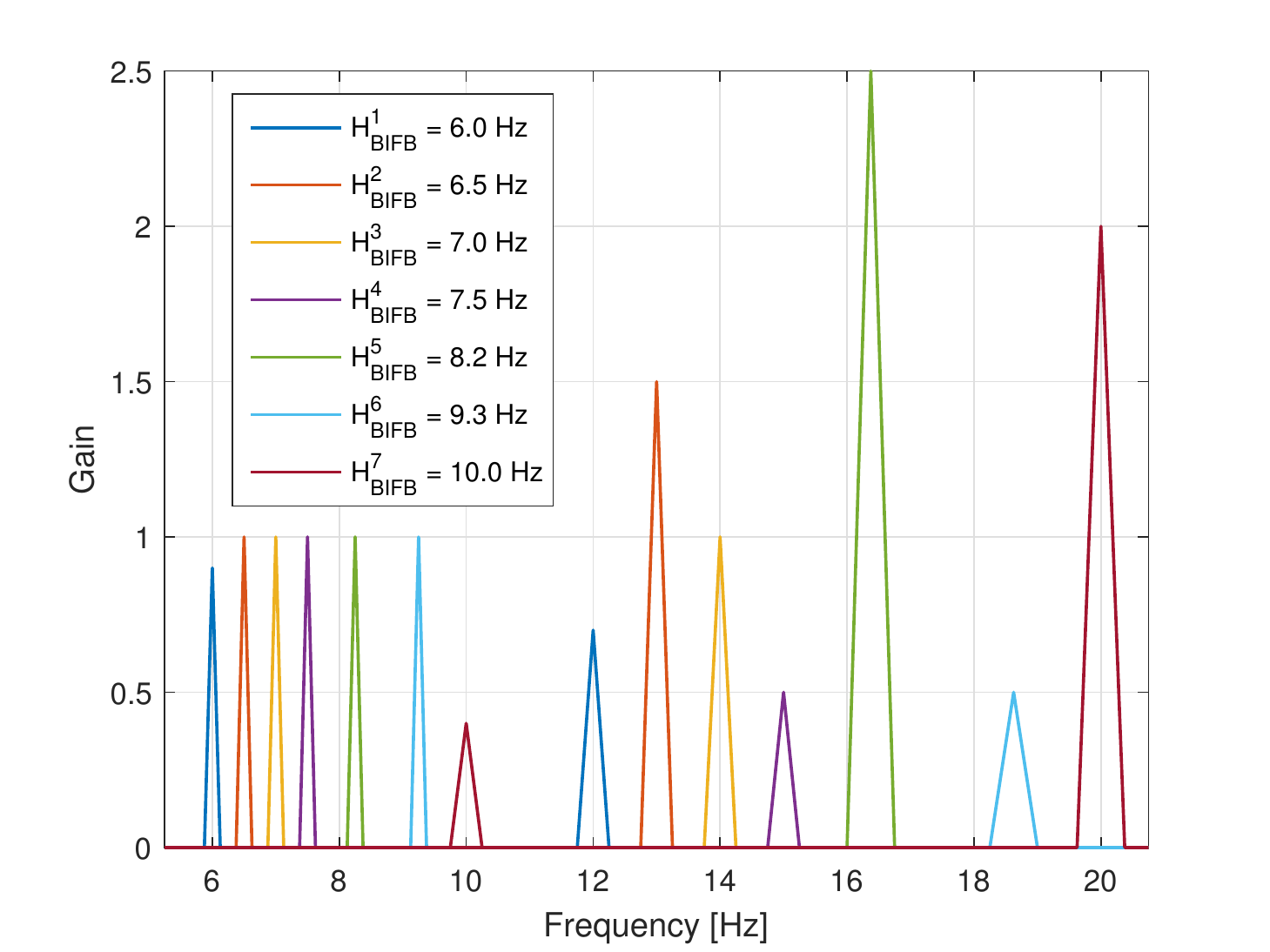}

}

\caption{(a) A sample BIFB design that mainly deals with low SNR at the high-frequency
band (Dataset-A); (b) A sample BIFB design that mainly deals with
frequency selectivity (Dataset-B). \label{fig:BIFB-Datasets}}
\end{figure*}

\begin{equation}
\begin{array}{cc}
h_{\theta}^{k}(\tilde{X})=g(\theta_{k}^{T}\tilde{X})=\frac{1}{1+e^{-\theta_{k}^{T}\tilde{X}}} & \forall k\end{array}\label{eq:HypFunc}
\end{equation}
where $g$ represents the sigmoid function, $\tilde{X}$ denotes the
augmented feature vector (i.e., $[1,x_{1},...x_{2K}]$) with a size
of $2K+1$, and $\theta_{k}$ stands for the mapping weight vector
of $k^{th}$ class. $\theta_{k}$ is chosen in such a way that it
minimizes the cost function $J(\theta_{k})$, which is a distance
metric between the prediction and the actual class label $(y)$, by
the following equation \cite{ngMLnotes}: 

\[
J(\theta_{k})=\frac{1}{M}\sum_{m=1}^{M}\left[-y^{(m)}\,log\left((h_{\theta}(\tilde{X}^{(m)})\right)-(1-y^{(m)})\right.
\]

\begin{equation}
\begin{array}{cc}
\left.\times\,log\left(1-h_{\theta}(\tilde{X}^{(m)})\right)\right]+{\displaystyle \frac{\lambda}{2M}}\sum_{j=1}^{2K}\theta_{k_{j}}^{2} & \forall k\end{array}\label{eq:CostFunc}
\end{equation}
where $\left\{ \left(X^{(m)},y^{(m)}\right);\:m=1,\text{\dots},M\right\} $
represents the training set with $M$ training examples and $y\:\epsilon\,\left\{ 0,1\right\} $.
The leave-one-out cross-validation is performed to resample the training
data for true objectivity and its suitability for small datasets \cite{alpaydin2012}.
The last summative term in Eq. \ref{eq:CostFunc} prevents over-fitting
the classifier and its precision is controlled by the regularization
parameter $\lambda$. $J(\theta_{k})$ is minimized with a gradient
descent algorithm, and optimal $\theta_{k}$ is calculated for $\forall k$.

\begin{figure*}[!b]
\centering\includegraphics[width=2\columnwidth]{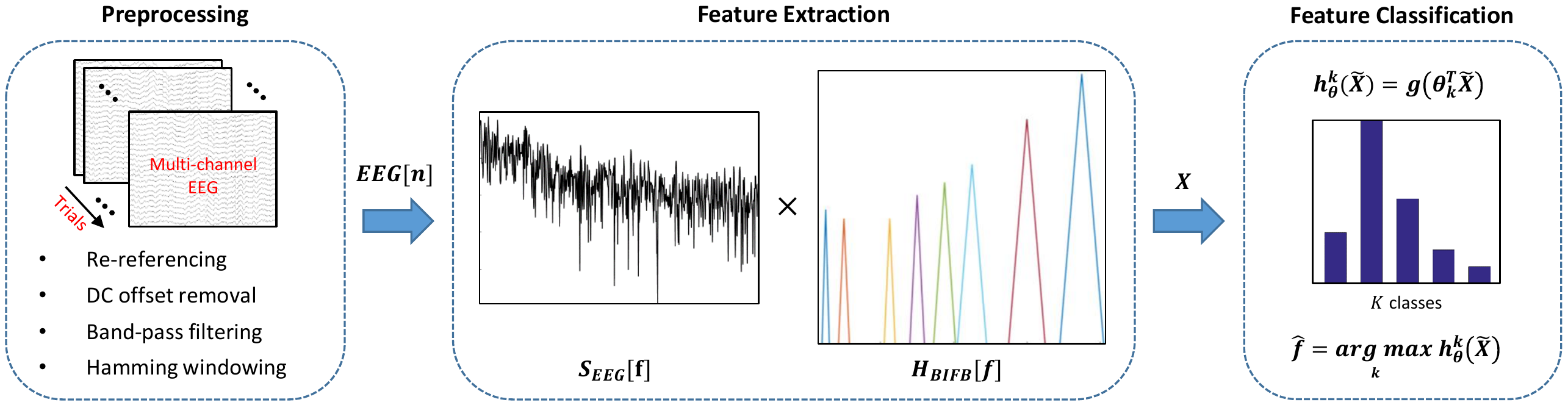}\caption{Flowchart of the signal processing stages of an SSVEP-based BCI using
the proposed BIFBs. \label{fig:FlowChart}}
\end{figure*}

After the training stage, the probability that a given input belongs
to each class is calculated using the hypothesis function in Eq. \ref{eq:HypFunc},
and the class with the highest probability is labeled as a candidate
frequency for recognition as follows:

\begin{equation}
f_{c}=\argmax_{k}h_{\theta}^{k}(X)\quad\forall k
\end{equation}
The candidate frequency is labeled as recognized (i.e., $\widehat{f}=f_{c}$)
when the same $f_{c}$ occurs at least $t$ times in the last $T$
iterations, where the typical values for these parameters are three
and four, respectively. If the selection criteria are not satisfied
during the given period, it is evaluated as an unsuccessful recognition.
A flowchart of the proposed BIFB method for SSVEP frequency recognition
is presented in Fig.~\ref{fig:FlowChart}.

\begin{table*}[t]
\caption{Performance Evaluation of SSVEP Recognition Algorithms on Dataset-A
\label{tab:Results-A}}

\centering\includegraphics[width=1.75\columnwidth]{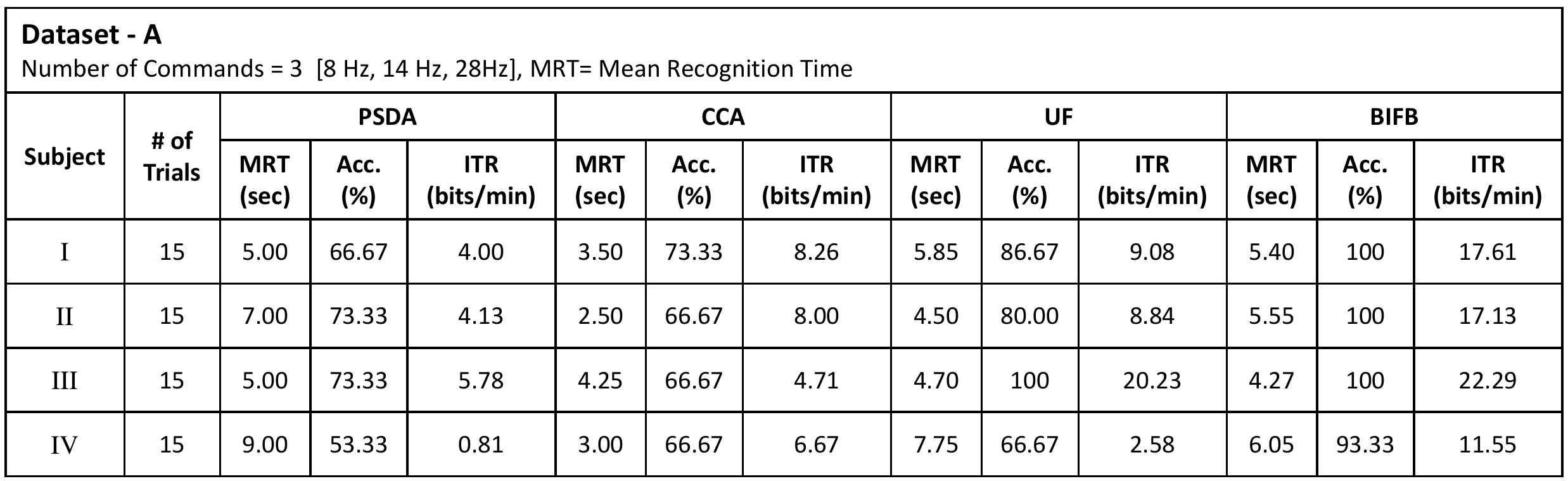}
\end{table*}

\begin{table*}[t]
\caption{Performance Evaluation of SSVEP Recognition Algorithms on Dataset-B\label{tab:Results-B}}

\centering\includegraphics[width=1.75\columnwidth]{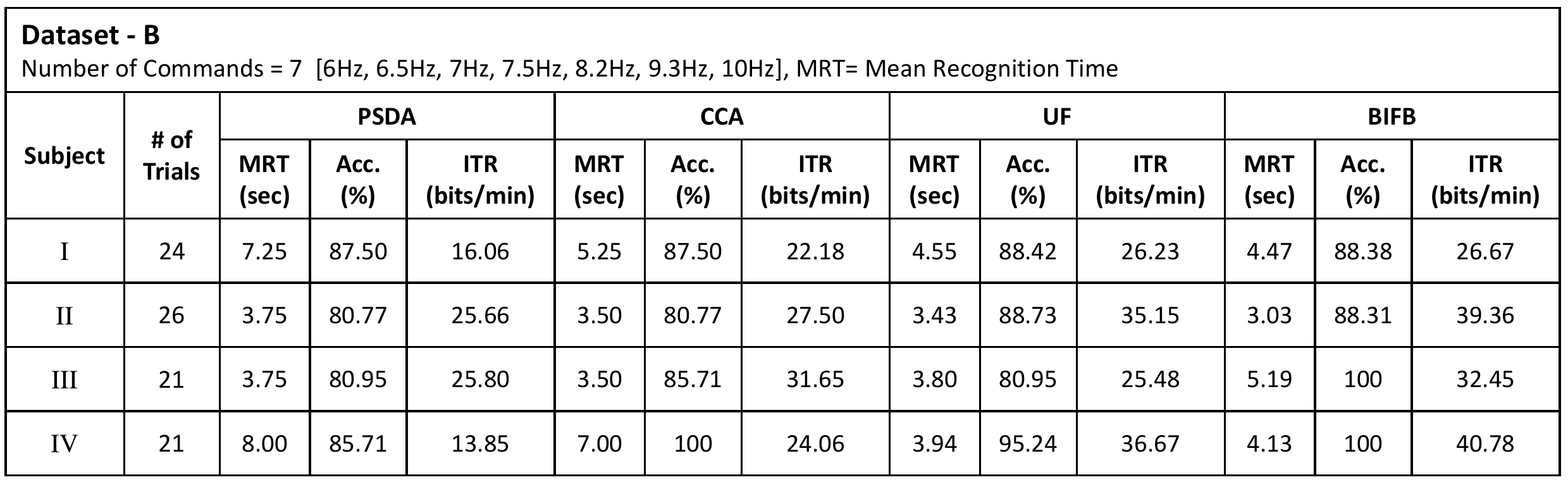}
\end{table*}

\subsection{Comparison Methods\label{subsec:Comparison-Methods}}

The performance of the proposed algorithm is compared with the performances
of various SSVEP frequency recognition algorithms. PSDA and CCA are
selected as comparison methods since they are the most common techniques
in the literature to compare a new algorithm \cite{zhang2013,chen2015,wang2016}.
However, there is no training in these traditional approaches, and
a direct comparison may not be proper. Therefore, the BIFBs are replaced
with unit filters (UFs), and a similar classical training process
is performed for classification to examine the effectiveness of the
proposed bio-inspired filter design fairly. Also, the parameters are
optimized/calibrated to maximize the ITR performance in all SSVEP
frequency recognition methods.

\medskip{}

\subsubsection{UF}

It is an SSVEP frequency recognition method, which follows a similar
procedure to the proposed scheme in Subsection \ref{subsec:PropAlg}
except for the utilization of BIFBs. Instead, the features are extracted
with unit filters, and they are expressed as follows: 

\begin{equation}
\begin{array}{cc}
H_{UF}^{k}[f]:\left\{ \begin{array}{cc}
1 & (\tilde{f_{k}}-BW_{D})\leq f\leq(\tilde{f_{k}}+BW_{D})\\
0 & otherwise
\end{array}\right. & \forall k\end{array}
\end{equation}
where $D$ is the index for dataset, and $BW_{D}$ equals to $1$
for Dataset-A whereas it is equal to $0.5$ for Dataset-B. Since the
only difference between BIFB and UF methods is the filter type utilized
in the feature extraction stage (like a controlled experiment), any
performance difference can be attributed to the filter bank design. 

\medskip{}

\subsubsection{PSDA}

The EEG signal from the occipital channel is pre-processed, and PSD
is estimated similar to the proposed approach. Afterward, the peak
of the spectrum is determined as the target frequency $(\widehat{f})$
in the traditional PSDA approach \cite{wang2006}. In this study,
the harmonic responses are considered in the PSDA algorithm as well
for a fair comparison. Initially, the class values, where each class
represents a target frequency, are calculated by summing the energy
in the fundamental frequency and harmonic bands. Subsequently, the
class that has the maximum value is recognized as SSVEP target frequency
as follows:

\begin{equation}
c_{k}={\displaystyle \sum_{f}{\textstyle S_{EEG}[f]\,H_{UF}^{k}[f]}}+{\displaystyle \sum_{f}{\textstyle S_{EEG}[f]\,H_{UF}^{K+k}[f]}}
\end{equation}

\begin{equation}
\widehat{f}=\max_{k}c_{k}\quad\forall k
\end{equation}

\subsubsection{CCA}

The final comparison method, CCA, is a multivariable statistical method
that aims to reveal the underlying correlation between two sets of
data \cite{hotelling1936} and has been widely used for SSVEP frequency
recognition \cite{lin2007}. If $A$ is a multi-channel EEG signal,
and $B$ is the Fourier series of a square-wave stimulus signal, CCA
searches for the linear combination vectors ($\gamma_{a}$, $\gamma_{b}$)
that maximize the correlation between $\alpha=\gamma_{a}^{T}A$ and
$\beta=\gamma_{b}^{T}B$ by optimizing the following equation:

\begin{equation}
\max_{\gamma_{a}\gamma_{b}}\rho(\alpha,\beta)=\frac{E[\gamma_{a}^{T}AB^{T}\gamma_{b}]}{\sqrt{E[\gamma_{a}^{T}AA^{T}\gamma_{a}]\,E[\gamma_{b}^{T}BB^{T}\gamma_{b}]}}\label{eq:CCA}
\end{equation}
The optimization problem in Eq. \ref{eq:CCA} can be solved by a generalized
eigenvalue decomposition \cite{friman2001}, and the maximum correlation
coefficient $(\rho)$ is computed for each $B_{k}$. Finally, the
SSVEP target frequency is recognized as follows:

\begin{equation}
\widehat{f}=\max_{k}\rho_{k}\quad\forall k
\end{equation}
A similar pre-processing procedure to PSDA is applied to the multi-channel
EEG signal (i.e., $A$) in CCA as well.

\begin{table}
\caption{Statistical Analysis of ITR Difference Between BIFB and Comparison
Methods by Using Paired t-Test\label{tab:Significance}}

\centering\includegraphics[width=0.6\columnwidth]{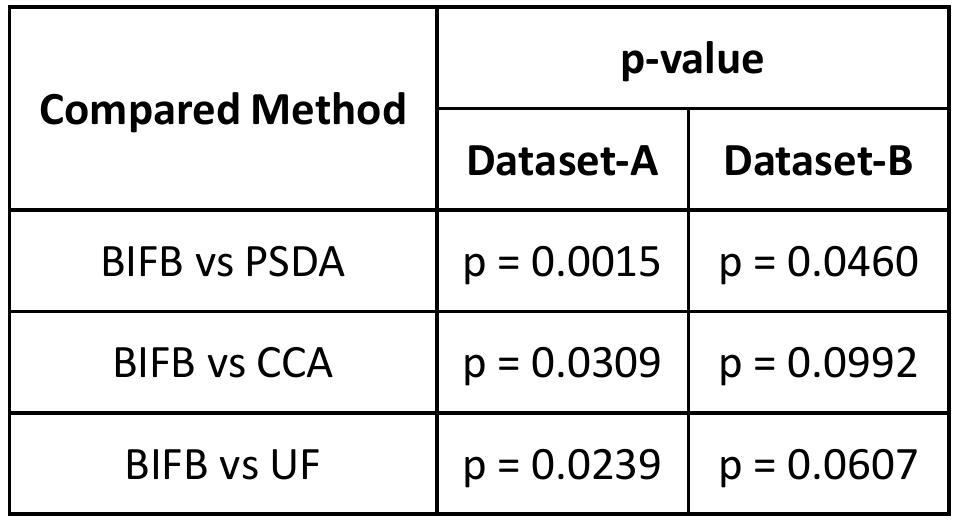}
\end{table}

\section{Results and Discussion\label{sec:III}}

The proposed BIFB method for SSVEP frequency recognition is tested
on two datasets that include EEG recordings of eight subjects in 152
trials. The system performance is evaluated in terms of mean recognition
time (MRT), recognition accuracy, and ITR by implementing a leave-one-out
cross-validation methodology. It is worth to note that ITR changes
logarithmically with the number of available commands in Eq.~\ref{eq:ITR}.
The number of commands in each dataset is different, and hence ITRs
need to be interpreted separately. The performance of the proposed
algorithm is compared with three baseline methods, and the results
are listed in Table \ref{tab:Results-A} and Table \ref{tab:Results-B}.
The statistical significance of these results is examined by paired
t-tests \cite{deGroot2012}, and corresponding p-values are presented
in Table~\ref{tab:Significance}. No multiple comparison correction
is considered since the study is restricted to a small number of planned
comparisons, and the results of individual tests are important \cite{armstrong2014}.

\begin{figure*}[t]
\subfloat[\label{fig:Accuracy}]{\includegraphics[width=1\columnwidth]{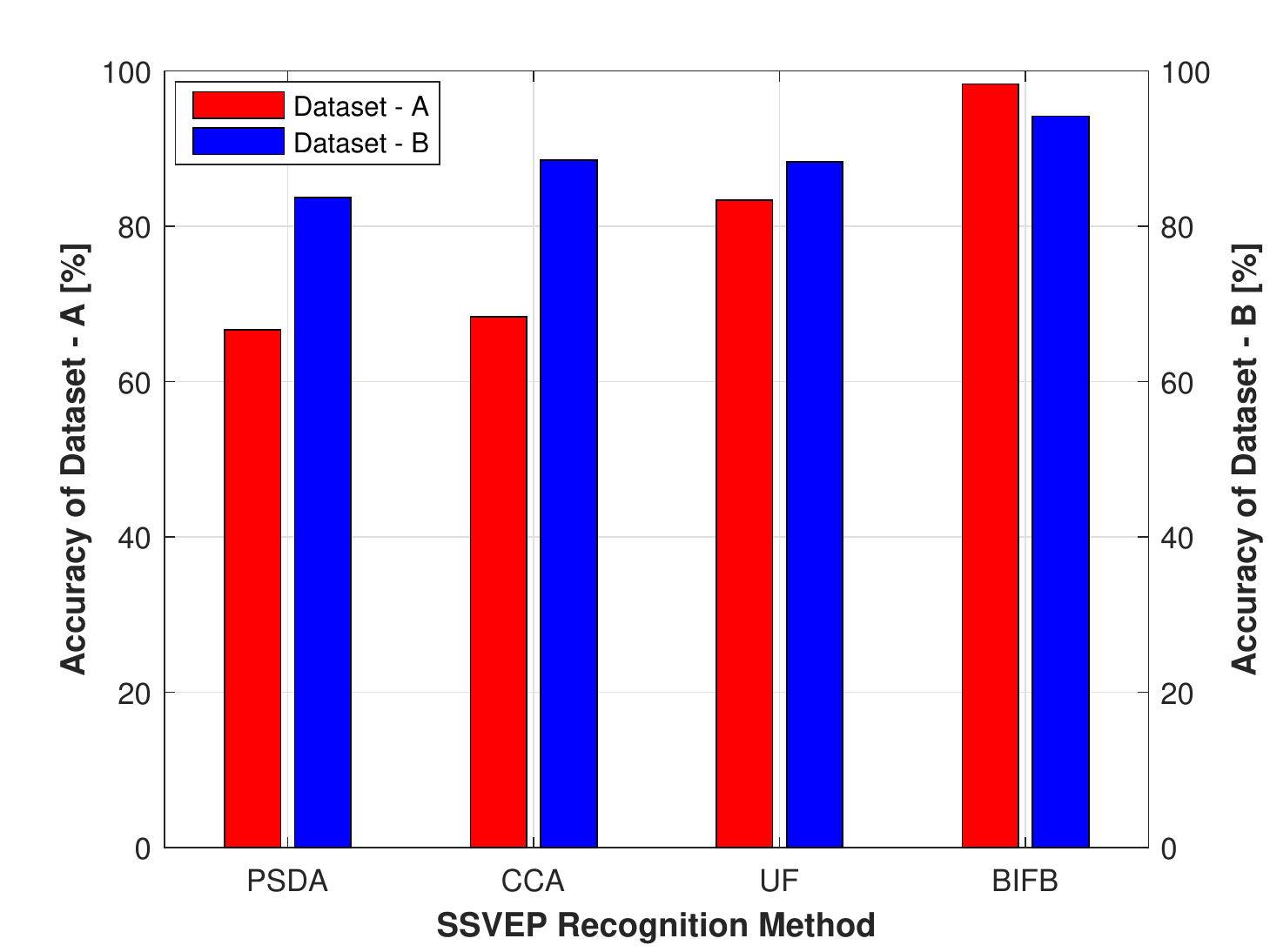}

}\subfloat[\label{fig:ITR}]{\includegraphics[width=1\columnwidth]{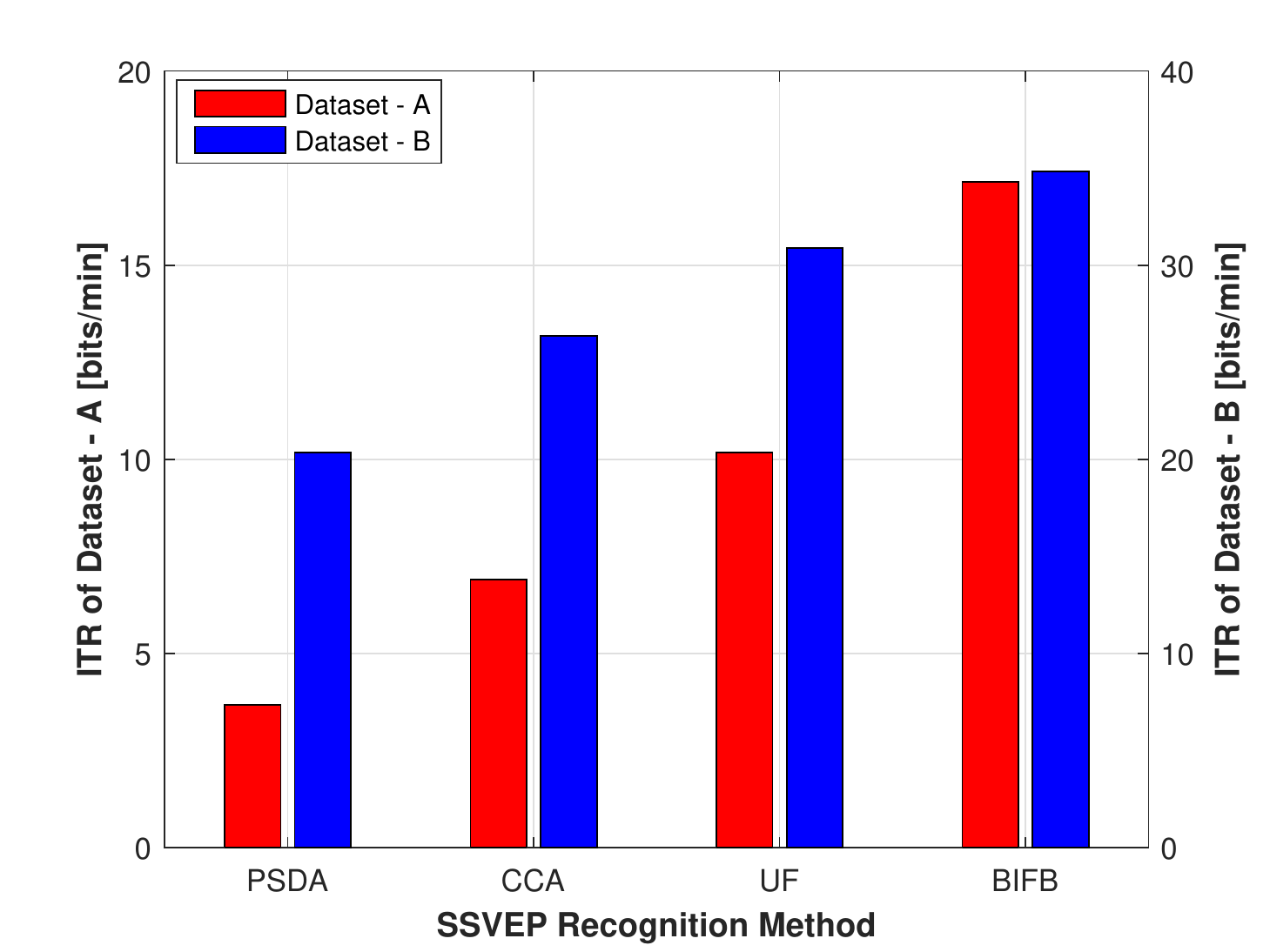}

}

\caption{The mean recognition accuracy and ITR performance of the SSVEP recognition
methods. \label{fig:ACCandITR}}
\end{figure*}

The traditional PSDA approach requires longer time windows compared
to the other three methods to provide sufficient accuracy, which leads
to a longer MRT and a lower ITR. A shorter MRT not only improves the
ITR but also diminishes the visual fatigue due to a reduced gazing
duration. Also, PSDA, as well as CCA, is incapable of detecting stimuli
in the high-frequency band. The low recognition accuracy of 28 Hz
stimulus, which is presented in Table \ref{tab:HighFreq}, explains
the poor performance results of these algorithms in Dataset-A. On
the other hand, there are no high-frequency stimuli in Dataset-B,
but the frequency selectivity decreases the ITR performances of PSDA,
CCA, and UF. 

PSDA and CCA have the advantage of not requiring training, and just
a straightforward calibration that includes the selection of electrode
locations, number of harmonics, and time window duration is sufficient
to perform the recognition. However, these algorithms disregard the
correlation information between the classes. A simple logistic regression
model can capture the between-class information and enhance performance.
Another classification model may achieve better performance. However,
it is beyond the scope of this study, and \cite{lotte2007,carvalho2015,oikonomou2016}
can be referred for more detailed information. The SSVEP response
is subject-specific, but the inter-trial variance is low within a
subject. Therefore, one-time individualized training is acceptable
to acquire a higher ITR. Furthermore, BIFB and UF implement the same
classifier. However, a feature extraction stage with BIFB, which captures
the underlying biological features of SSVEPs, increases the separability
and outperforms UF for SSVEP frequency recognition in both datasets.

\begin{table}[t]
\caption{SSVEP recognition accuracy performance for 28 Hz Stimulus in Dataset-A
\label{tab:HighFreq}}

\centering\includegraphics[width=0.35\columnwidth]{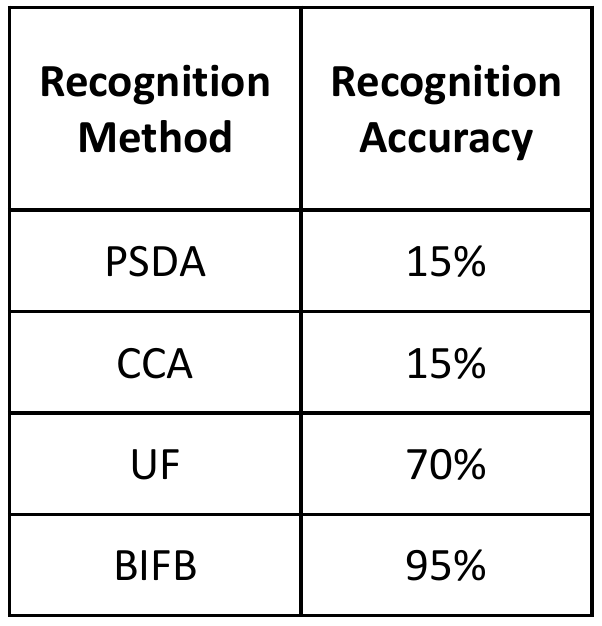}
\end{table}

User comfort is another important criterion in BCI design besides
the ITR. It is reported that high-frequencies cause less visual fatigue
induced by the flicker \cite{wang2006,diez2011}. The promising performance
of BIFBs in the high-frequency band may let the designers include
this low SNR band in their BCI system. As a result, the user discomfort
caused by the flicker reduces, and also ITR increases due to the increase
in number of available commands. Furthermore, the number of electrodes
is critical for user comfort. Although it is preferable to have a
dense sensor system while mapping the brain, it is not suitable for
practical BCI applications. In this study, BIFB utilized the information
from one electrode for the sake of simplicity. The results show that
a single-channel algorithm can provide superior performance compared
to a multi-channel algorithm (i.e., CCA), and enhance user comfort
as well. However, the use of BIFBs is not restricted to single-channel
utilization, and recognition accuracy might be further improved by
taking advantage of multi-channel information in the feature extraction
stage. For example, a simple way to utilize the BIFBs with multi-channel
EEG would be to apply them on signals from the occipital channels
and pass the weighted average of the extracted features to the feature
classification stage.

\section{Conclusions\label{sec:IV}}

A novel SSVEP recognition method that exploits the inherent biological
characteristics of SSVEPs is introduced in this paper. The BIFBs capture
frequency selectivity, subject specificity, and harmonic SSVEP responses
in the feature extraction stage and enhance the separability of classes.
The proposed method is tested on two benchmark datasets available
online and outperforms several recognized recognition algorithms.
The BIFBs are promising particularly in the high-frequency band where
SNR is low. Hence, this method not only increases the ITR of an SSVEP
based BCI but also might improve its user comfort due to less visual
fatigue. The results show the potential of bio-inspired design, and
the findings will be extended to include further SSVEP characteristics.
First, the best pulse shape to utilize in the filter banks remains
unknown. The triangular filters in this study might need to be replaced
with another shape such as Gaussian or raised-cosine to improve the
performance further. Second, the BIFBs should incorporate the time-characteristics
of SSVEPs. The onset-delay of the response is frequency selective
\cite{bakardjian2010} and including this distinct feature might increase
the recognition accuracy as well. Last, the SSVEP response also strongly
depends on the stimuli type \cite{zhu2010,teng2012}, and the BIFB
adaptation considering the visual stimuli requires further investigation.

BCIs and their associated technologies will shape the future of communication,
control, and security as a part of WBAN. To fully exploit and further
increase the potential of these devices, it is necessary to employ
an accurate model of the driving physiological signal in the recognition
algorithm. Bio-inspired designs such as the proposed BIFBs will be
the key in enabling the development of reliable, efficient, and high-performance
BCI systems.

\vspace{0.5mm}

\bibliographystyle{IEEEtran}
\bibliography{IEEEabrv,BIFB_Ref}
\vspace{-11mm}
\begin{IEEEbiography}[\includegraphics{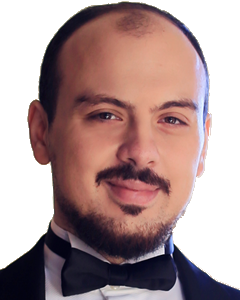}]{Ali Fatih Demir}
(S'08) received the B.S. degree in electrical engineering from Y\i ld\i z
Technical University, Istanbul, Turkey, in 2011 and the M.S. degrees
in electrical engineering and applied statistics from Syracuse University,
Syracuse, NY, USA in 2013. He is currently pursuing the Ph.D. degree
as a member of the Wireless Communication and Signal Processing (WCSP)
Group in the Department of Electrical Engineering, University of South
Florida, Tampa, FL, USA. His current research interests include PHY
and MAC aspects of wireless communication systems, \emph{in vivo}
wireless communication systems, and signal processing/machine learning
algorithms for brain-computer interfaces.

\vspace{-9mm}
\end{IEEEbiography}

\begin{IEEEbiography}[\includegraphics{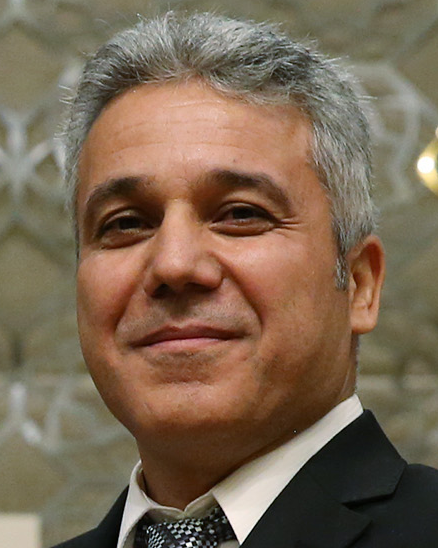}]{Huseyin Arslan}
 (S\textquoteright 95\textendash M\textquoteright 98\textendash SM\textquoteright 04\textendash F\textquoteright 16)
received the B.S. degree in electrical and electronics engineering
from Middle East Technical University, Ankara, Turkey, in 1992, and
the M.S. and Ph.D. degrees in electrical engineering from Southern
Methodist University, Dallas, TX, USA, in 1994 and 1998, respectively.
From January 1998 to August 2002, he was with the research group of
Ericsson Inc., NC, USA, where he was involved with 2G and 3G wireless
communication systems. He is currently a Professor of Electrical Engineering
at the University of South Florida, Tampa, FL, USA, and the Dean of
the College of Engineering and Natural Sciences at the \.{I}stanbul
Medipol University, \.{I}stanbul, Turkey. His current research interests
are on 5G and beyond, waveform design, advanced multiple accessing
techniques, physical layer security, beamforming and massive MIMO,
cognitive radio, dynamic spectrum access, interference management
(avoidance, awareness, and cancellation), co-existence issues on heterogeneous
networks, aeronautical (high altitude platform) communications, millimeter-wave
communications and\emph{ in vivo} communications. He is currently
a member of the editorial board for the \emph{IEEE Communications
Surveys and Tutorials} and the \emph{Sensors Journal}.

\vspace{-12cm}
\end{IEEEbiography}

\begin{IEEEbiography}[\includegraphics{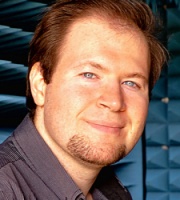}]{Ismail Uysal}
 (S\textquoteright 04\textendash M\textquoteright 08) received the
B.S. degree in electrical and electronics engineering from Middle
East Technical University, Ankara, Turkey in 1998, and the M.S. and
Ph.D. degrees in electrical and computer engineering from the University
of Florida (UF), Gainesville, FL, USA in 2006 and 2008. From 2008
to 2010, he was a postdoctoral research fellow at the UF Research
Center for Food Distribution and Retailing. Since 2010, he has been
with the University of South Florida where he is currently an assistant
professor of electrical engineering and the director of the radio
frequency identification (RFID) Lab for Applied Research under the
College of Engineering. His research interests include deep machine
learning theory and applications in semi-supervised and unsupervised
settings, data-oriented applications of RFID systems in healthcare
and food supply chains, and signal processing algorithms for brain-computer
interfaces.
\end{IEEEbiography}

\end{document}